\documentclass[12pt]{article}
\usepackage[cp1251]{inputenc}
\usepackage[russian]{babel}
\begin{document}
\large
\par
\begin{center}
{\bf Remarks to the Standard Theory of Neutrino Oscillations.
\par
Alternative Scheme of Neutrino Oscillations}  \\
\par
Beshtoev Kh. M.
\par
Joint Institute for Nuclear Research, Joliot Curie 6, 141980
Dubna, Moscow region, Russia \\
\end{center}

\par
{\bf Abstract} \\

\par
In the standard theory of neutrino oscillations it is supposed
that physical observed neutrino states $\nu_{e}, \nu_{\mu },
\nu_{\tau}$ have no definite masses, that they are initially
produced as a mixture of the $\nu_{1}, \nu_{2}, \nu_{3}$ neutrino
states (are produced as a wave packet), and that neutrino
oscillations are the real ones. Then this wave packet must
decompose at a definite distance into constituent parts and
neutrino oscillations must disappear. It was shown that these
suppositions lead to violation of the law of energy and momentum
conservation. An alternative scheme of neutrino oscillations
obtained within the framework of particle physics has been
considered where the above mentioned shortcomings are absent, the
oscillations of neutrinos with equal masses are the real ones, and
the oscillations of neutrinos with different masses are virtual
ones. Expressions for probabilities of neutrino transitions
(oscillations) in the
alternative (corrected) scheme, are given. \\

\par
{\bf 1. Introduction} \\

\par
The suggestion that, in analogy with $K^{o},\bar K^{o}$
oscillations, there could be neutrino-antineutrino oscillations (
$\nu \rightarrow \bar \nu$) was considered by Pontecorvo [1] in
1957. It was subsequently considered by Maki et al. [2] and
Pontecorvo [3] that there could be mixings (and oscillations) of
neutrinos of different flavors (i.e., $\nu _{e} \rightarrow \nu
_{\mu }$ transitions). In the standard theory of neutrino
oscillations [4] it is supposed that physical observed neutrino
states $\nu_{e}, \nu_{\mu }, \nu_{\tau}$ have no definite masses
and that they are directly produced as mixture of the $\nu_{1},
\nu_{2}, \nu_{3}$ neutrino states (as wave packets). Below we
discuss the consequences of these suppositions and then an
alternative scheme of neutrino oscillations constructed in the
framework of particle physics theory is considered. This work is
expanded version of the previous work hep-ph/0503202.
\par
Come to consideration of basic elements and shortcomings of the
standard theory of neutrino oscillations. \\

\par
{\bf 2. Basic Elements of the Standard Theory of Neutrino
Oscillations}
\\

\par
In the standard theory of neutrino oscillations [4], constructed
in the framework of Quantum theory (Mechanics) in analogy with the
theory of $K^{o}, \bar{K}^{o}$ oscillation, it is supposed that
mass eigenstates are $\nu_{1}, \nu_{2}, \nu_{3}$ neutrino states
but not physical observed neutrino states $\nu_{e}, \nu_{\mu },
\nu_{\tau}$. And that the neutrinos $\nu_{e}, \nu_{\mu },
\nu_{\tau}$ are directly produced as superpositions of $\nu_{1},
\nu_{2}, \nu_{3}$ states (neutrinos). Since $\nu_{e}, \nu_{\mu },
\nu_{\tau}$ neutrinos are superpositions of $\nu_{1}, \nu_{2},
\nu_{3}$ neutrinos (mass matrix is nondiagonal) then they have no
definite mass. And also that neutrino oscillations are real
oscillations even, i.e. that there is a real transition of
electron neutrino $\nu_e$ into muon neutrino $\nu_{\mu}$ (or tau
neutrino $\nu_{\tau}$). Come to consideration of oscillations in
the framework of this approach. To simplify, the case of two
neutrinos is considered.
\par
The mass lagrangian of two neutrinos ($\nu_e, \nu_\mu$) has the
following form:
$$
\begin{array}{c}{\cal L}_{M} = - \frac{1}{2} \left[m_{\nu_e}
\bar \nu_e \nu_e + m_{\nu_\mu} \bar \nu_{\mu} \nu_{\mu } +
m_{\nu_e \nu_{\mu }}(\bar \nu_e \nu_{\mu } + \bar \nu_{\mu }
\nu _e) \right] \equiv \\
\equiv  - \frac{1}{2} (\bar \nu_e, \bar \nu_\mu)
\left(\begin{array}{cc} m_{\nu_e} & m_{\nu_e \nu_{\mu }} \\
m_{\nu_{\mu} \nu_e} & m_{\nu_\mu} \end{array} \right)
\left(\begin{array}{c} \nu_e \\ \nu_{\mu } \end{array} \right)
\end{array} ,
\eqno(1)
$$
which is diagonalized by rotation on the angle $\theta$ and then
this lagrangian (1) transforms into the following one (see ref. in
[4]):
$$
{\cal L}_{M} = - \frac{1}{2} \left[ m_{1} \bar \nu_{1} \nu_{1} +
m_{2} \bar \nu_{2} \nu_{2} \right]  , \eqno(2)
$$
where
$$
m_{1, 2} = {1\over 2} \left[ (m_{\nu_e} + m_{\nu_\mu}) \pm
\left((m_{\nu_e} - m_{\nu_\mu})^2 + 4 m^{2}_{\nu_\mu \nu_e}
\right)^{1/2} \right] ,
$$
\par
\noindent and angle $\theta $ is determined by the following
expression:
$$
tg 2 \theta  = \frac{2 m_{\nu_e \nu_\mu}} {(m_{\nu_\mu} -
m_{\nu_e})} , \eqno(3)
$$
$$
\begin{array}{c}
\nu_e = cos \theta  \nu_{1} + sin \theta \nu_{2}  ,         \\
\nu _{\mu } = - sin \theta  \nu_{1} + cos \theta  \nu_{2} .
\end{array}
\eqno(4)
$$
From eq.(3) one can see that if $m_{\nu_e} = m_{\nu_{\mu}}$, then
the mixing angle is equal to $\pi /4$ independently of the value
of $m_{\nu_e \nu_\mu}$:
\par
The expression for time evolution of $\nu _{1}, \nu _{2}$
neutrinos (see (2), (4)) with masses $m_{1}$ and $m_{2}$ is
\par
$$
\nu _{1}(t) = e^{-i E_1 t} \nu _{1}(0),  \qquad \nu _{2}(t) =
e^{-i E_2 t} \nu _{2}(0) , \eqno(5)
$$
where
$$
E^2_{k} = (p^{2} + m^2_{k}), k = 1, 2 .
$$
\par
If neutrinos are propagating without interactions, then
\par
$$
\begin{array}{c}
\nu_e(t) = cos \theta e^{-i E_1 t} \nu_{1}(0) + sin \theta
e^{-i E_2 t} \nu_{2}(0) , \\
\nu_{\mu }(t) = - sin \theta e^{-i E_1 t} \nu_{1}(0) + cos \theta
e^{-i E_2 t} \nu_{2}(0) .
\end{array}
\eqno(6)
$$
\noindent Using the expression for $\nu _{1}$ and $\nu _{2}$  from
(11), and putting it into (6), one can get the following
expression:
$$
\nu_e (t) = \left[e^{-i E_1 t} cos^{2} \theta + e^{-i E_2 t}
sin^{2} \theta \right] \nu _e (0) +
$$
$$
+ \left[e^{-i E_1 t} - e^{-i E_2 t} \right] sin \theta \cos \theta
\nu_{\mu }(0) , \eqno(7)
$$
$$
\nu_{\mu }(t) = \left[e^{-i E_1 t} sin^{2} \theta + e^{-i E_2 t}
cos^{2} \theta \right] \nu_{\mu}(0)  +
$$
$$
+ \left[e^{-i E_1 t} - e^{-i E_2 t} \right] sin\theta cos \theta
\nu_e (0) .
$$
\par
The probability that neutrino $\nu_e$ produced at the time $t = 0$
will be transformed into $\nu_{\mu}$ at the time $t$ is an
absolute value of amplitude $\nu_{\mu}(0)$ in (7) squared, i. e.
\par
$$
\begin{array}{c}
P(\nu_e \rightarrow \nu_{\mu}) = \mid(\nu_{\mu}(0) \cdot \nu_e(t)) \mid^2 =\\
 = {1\over 2} \sin^{2} 2\theta \left[1 - cos ((m^{2}_{2} - m^{2}_{1}) / 2p)
t \right] ,
\end{array}
\eqno(8)
$$
\noindent
where it is supposed that $p \gg  m_{1}, m_{2}; E_{k}
\simeq p + m^{2}_{k} / 2p$.
\par
Besides, since $\nu_e, \nu_{\mu}, \nu_{\tau}$ neutrinos are
superpositions of $\nu_{1}, \nu_{2}, \nu_{3}$, then the $\nu_e,
\nu_{\mu}, \nu_{\tau}$ neutrinos are wave packets having widths.
Then these $\nu_e, \nu_{\mu}, \nu_{\tau}$ states (neutrinos) are
unstable ones and must decompose for the time $\Delta t$ which is
determinated by the uncertainty relation [5, 6],
$$
\Delta t \sim \frac{L_{cohe}}{c}, \eqno(9)
$$
$$
L_{cohe} \cong \frac{4 E^2_{\nu} \Delta x}{\Delta m^2},
$$
where $c$ is the light velocity, $\Delta x$ is size of the object
where the physical observed neutrino is produced, $\Delta m^2$ is
squared neutrino mass differences ($\Delta m^2 \to m^2_{\nu_2} -
m^2_{\nu_1}$ or $m^2_{\nu_3} - m^2_{\nu_1}$). \\

\par
{\bf 2. Remarks to the Standard Theory of Neutrino Oscillations}
\\

\par
Now it is necessary to check: is it possible to prove main
suppositions of the standard theory of neutrino oscillations
within the framework of the particle physics theory (or the
relativistic quantum theory)?
\par
1. The mass eigenstates are $\nu_{1}, \nu_{2}, \nu_{3}$ neutrino
states but not physical observed neutrino states $\nu_{e},
\nu_{\mu }, \nu_{\tau}$. And then the neutrinos $\nu_{e}, \nu_{\mu
}, \nu_{\tau}$ are directly produced as superpositions of the
$\nu_{1}, \nu_{2}, \nu_{3}$ states (neutrinos).
\par
This supposition violates the causality principle since at
productions of $\nu_{e}, \nu_\mu, \nu_{\tau}$ neutrinos they
already know that they must be superpositions of the $\nu_{1},
\nu_{2}, \nu_{3} $ neutrinos.
\par
One of the basic positions of the particle physics theory (or the
quantum theory) [5, 7] is that particles must be produced in
eigenstates, i.e., particles are produced in states with a
diagonal mass matrix. For example, we have two interactions:
interaction with the lepton number conservations (interaction with
$W, Z$ exchanges) and interaction with the lepton number
violations (hypothetical interaction which is described by the
nondiagonal terms of the Cabibbo-Kobayashi-Maskawa matrices). What
states will be produced? It is clear that in the first case the
$\nu_{e}, \nu_{\mu }, \nu_{\tau}$ neutrinos will be produced and
in the second case the $\nu_1, \nu_2, \nu_3$ neutrinos will be
produced since they are eigenstates of the corresponding
interactions. Why are the $\nu_{e}, \nu_{\mu }, \nu_{\tau}$
neutrinos produced but we do not observe $\nu_1, \nu_2, \nu_3$
neutrino productions? Within the framework of the particle physics
theory it is possible only if the interaction with lepton number
violations has time to produce the $\nu_1, \nu_2, \nu_3$
neutrinos, i.e., we do not observe productions of these neutrinos
since the probabilities of their productions are very small [8].
Then, after productions of the $\nu_{e}, \nu_{\mu }, \nu_{\tau}$
neutrinos, since we cannot switch off the weak interaction which
violated the lepton numbers, they will be transformed to
superpositions of the $\nu_1, \nu_2, \nu_3$ neutrinos. So, one can
see that within the framework of the particle physics theory there
is no possibility for direct productions of particles in
superposition states.
\par
The same situation takes place in the hadron case, when in the
strong interactions (where strangeness is conserved) $K^o, \bar
K^o$ mesons (eigenstates) are produced. And then by the weak
interactions (where strangeness is violated) they are transformed
to superpositions of the $K^o_1, K^o_2$ mesons (eigenstates of the
weak interactions) and then oscillations take place [8, 9].
\par
Now let us discuss other consequences of the standard theory of
neutrino oscillations.
\par
2. Since the $\nu_{e}, \nu_{\mu }, \nu_{\tau}$ neutrinos are
directly produced as superpositions of the $\nu_{1}, \nu_{2},
\nu_{3}$ neutrinos (their mass matrix is nondiagonal), they cannot
have definite masses. Only $\nu_{1}, \nu_{2}, \nu_{3}$ neutrinos
have definite masses.
\par
As a consequence of these suppositions, we cannot formulate the
law of energy-momentum conservation in a strict form in the
processes with participation of these neutrinos.
\par
And it is also supposed that oscillations between the $\nu_{e},
\nu_{\mu }, \nu_{\tau}$ neutrinos are real oscillations.
\par
However, computation with (1)-(4) has shown that $\nu_e, \nu_\mu$
masses are
$$
m_{\nu_e} = m_1 cos^2 \theta + m_2 sin^2 \theta ,
$$
$$
m_{\nu_\mu} = m_1 sin^2 \theta + m_2 cos^2 \theta , \eqno(10)
$$
i.e., the $\nu_e, \nu_\mu$ neutrinos have definite masses which
are expressed via the $\nu_1, \nu_2$ masses and the mixing angle
$\theta$. It means that the supposition that the $\nu_e, \nu_\mu$
neutrinos have no definite masses is not confirmed. Then, if
neutrino oscillations are real oscillations, i.e. there is a real
transition of the electron neutrino $\nu_e$ into the muon neutrino
$\nu_{\mu}$ (or tau neutrino-$\nu_{\tau}$), the neutrino $x = \mu,
\tau$ will decay to an electron neutrino plus something:
$$
\nu_{x} \rightarrow \nu_e + ... \quad  . \eqno(11)
$$
As a result, we can get energy from vacuum, which is equal to the
mass difference (if $m_{\nu_x} > m_{\nu_e}$)
$$
\Delta E \sim m_{\nu_{x}} - m_{\nu_e} . \eqno(12)
$$
Then, again, this electron neutrino is converted into the muon
neutrino, which decays again and we get energy, etc. {\bf So we
have got a perpetuum mobile!} Obviously, the law of energy and
momentum conservation in these processes is not fulfilled.
\par
It is necessary to stress that these suppositions are in
contradiction with the fundamental demand of the particle physics
theory that the particles must have definite masses and the law of
energy-momentum conservation must be fulfilled in processes.
\par
3. The $\nu_{e}, \nu_{\mu }, \nu_{\tau}$ neutrinos are
superpositions of the $\nu_{1}, \nu_{2}, \nu_{3}$ neutrinos and
they are produced as wave packets and must decompose, i.e., at
distances $L$ when
$$
L > L_{cohe} , \eqno(13)
$$
from the point of their productions the wave packets decompose to
components and neutrino oscillations will be absent. Then we must
see $\nu_{1}, \nu_{2}, \nu_{3}$ neutrino states but not the states
of $\nu_{e}, \nu_{\mu }, \nu_{\tau}$ neutrinos. Neutrinos are
elementary particles. Within the framework of the elementary
particle theory the particles are produced as individual
eigenstates of the corresponding interaction. We can construct a
wave packet as superposition of individual particles having a
definite width only after their productions, but we cannot produce
a wave packet as an elementary particle within the framework of
the elementary particle theory (or the quantum theory).
\par
It also means that the Solar neutrinos cannot reach the Earth as
$\nu_e, \nu_{\mu}, \nu_{\tau}$ neutrino states.
$$
L_{cohe} \sim \frac{4 E^2_{\nu} \Delta x}{\Delta m^2} = 2.2 \cdot
10^{6} cm , \eqno(14)
$$
where $E=7 MeV$, $\Delta m^2 = 8.9 \cdot 10^{-5} eV^2$, $\Delta x
= 10^{-12} cm$ (the neutrinos are produced inside the nucleus).
However in experiments [10, 11] we see namely $\nu_e, \nu_\mu,
\nu_\tau$ neutrino states but not $\nu_{1}, \nu_{2}, \nu_{3}$
neutrino states.
\par
Without any doubt this standard theory requires a correction in
order to get rid of the above mentioned defects. Below we come to
construction of a correct scheme within the framework of the
elementary particle theory (or the quantum theory). \\

\par
{\bf 3. Alternative Scheme of Neutrino Oscillations} \\

\par
In the framework of the  particle physics theory [7] all particles
are stable ones or if they have widths then they must decay in the
states (particles) with small masses. It is a requirement which
must be fulfilled in the framework of particle physics theory. If
particles are wave packets then these wave packets will decompose
and we cannot obtain stable long-life particles.
\par
The only way to restore the law of energy-momentum conservation in
processes of neutrino oscillations  is to work in the framework of
particle physics theory. Then, these oscillations will be virtual
if neutrinos have different masses and these oscillations will
proceed in the framework of the uncertainty relations.
\par
So, the correct theory of neutrino oscillations can be constructed
only into the framework of the particle physics theory, where the
conception of mass shell is present [7, 12]. Besides, every
particle must be produced on its mass shell and it will be left on
its mass shell while passing through vacuum.
\par
In the considered scheme of neutrino oscillations, constructed in
the framework of the particle physics theory, it is supposed
(according to the experiments) that:
\par
1)  The  physical  observable neutrino states $\nu_{e}, \nu_{\mu
}, \nu_{\tau}$ are eigenstates of the weak interaction with $W,
Z^o$ exchanges. And, naturally, the mass matrix of $\nu_{e},
\nu_{\mu }, \nu_{\tau}$ neutrinos is diagonal, i.e., the mass
matrix of $\nu_e, \nu_\mu$ and $\nu_\mu$ neutrinos has the
following diagonal form (since these neutrinos are produced in the
weak interactions, it means that they are eigenstates of these
interactions and their mass matrix must be diagonal):
$$
\left(\begin{array}{ccc} m_{\nu_e}& 0& 0 \\ 0 & m_{\nu_\mu} & 0\\
0 & 0 & m_{\nu_\tau} \end{array} \right) . \eqno(15)
$$
Besides, all  the  available, experimental results indicate that
the lepton numbers $l_{e}, l_{\mu }, l_{\tau}$  are   well
conserved, i.e. the standard weak interactions (with $W, Z^o$
bosons) do not violate the lepton numbers.
\par
2) Then, to violate the  lepton  numbers, it  is  necessary  to
introduce an interaction violating these numbers. It is equivalent
to introducing of the nondiagonal  mass terms  in the  mass matrix
of $\nu_{e}, \nu_{\mu }, \nu_{\tau}$ neutrinos:
$$
M(\nu_e, \nu_\mu, \nu_\tau) = \left(\begin{array}{ccc} m_{\nu_e}&
m_{\nu_e \nu_\mu} & m_{\nu_e \nu_\tau}
\\ m_{\nu_\mu \nu_e} & m_{\nu_\mu} & m_{\nu_\mu \nu_\tau} \\
m_{\nu_\tau \nu_e} & m_{\nu_\tau \nu_\mu} & m_{\nu_\tau}
\end{array} \right) . \eqno(16)
$$
Diagonalizing this matrix [4]
$$
M(\nu_e, \nu_\mu, \nu_\tau) = V^{-1} M(\nu_1, \nu_2, \nu_2) V ,
\eqno(17)
$$
we go to the $\nu_{1}, \nu _{2}, \nu_{3}$ neutrino mass matrix
$$
\left(\begin{array}{ccc} m_{\nu_1}& 0& 0 \\ 0 & m_{\nu_2} & 0\\
0 & 0 & m_{\nu_3} \end{array} \right) , \eqno(18)
$$
where $V$ is neutrino mixings matrix $V$. Then the vector state
$\Psi(\nu_e, \nu_\mu, \nu_\tau)$, of $\nu_e, \nu_\mu, \nu_\tau$
neutrinos
$$
\Psi(\nu_e, \nu_\mu, \nu_\tau) = \left(\begin{array}{c} \nu_e \\
\nu_\mu \\ \nu_\tau
\end{array} \right) , \eqno(19)
$$
is transformed into the vector state $\Psi(\nu_1, \nu_2, \nu_2)$
of $\nu_1, \nu_2, \nu_2$ neutrinos
$$
\Psi(\nu_e, \nu_\mu, \nu_\tau) = V \Psi(\nu_1, \nu_2, \nu_2) ,
\eqno(20)
$$
i.e., $\nu_e, \nu_\mu, \nu_\tau$ neutrinos are transformed into
superpositions of $\nu_1, \nu_2, \nu_2$ neutrinos.
\par
We can choose parameterization of this matrix $V$ in the form
proposed by Maiani [13] $V$, then
$$
{V = \left( \begin{array} {ccc}1& 0 & 0 \\
0 & c_{\gamma} & s_{\gamma} \\ 0 & -s_{\gamma} & c_{\gamma} \\
\end{array} \right) \left( \begin{array}{ccc} c_{\beta} & 0 &
s_{\beta} \exp(-i\delta) \\ 0 & 1 & 0 \\ -s_{\beta} \exp(i\delta)
& 0 & c_{\beta} \end{array} \right) \left( \begin{array}{ccc}
c_{\theta} & s_{\theta} & 0 \\ -s_{\theta} & c_{\theta} & 0 \\ 0 &
0 & 1 \end{array}\right)} , \eqno(21)
$$
where $\theta, \beta, \gamma$ and $\delta$ are angles of neutrino
mixings and parameter of $CP$ violation.
\par
Exactly like the case  of $K^{o}$ mesons produced  in strong
interactions, when mainly $K^{o}, \bar{K}^{o}$ mesons are produced
but not $K_1, K_2$ mesons. In  the considered case $\nu_{e},
\nu_{\mu }, \nu_{\tau}$, but not $\nu_{1}, \nu_{2}, \nu_{3}$,
neutrino  states are mainly produced in the weak interactions
(this is so since the contribution of the lepton numbers violating
interactions  in this process is too small).
\par
3) Then, when the $\nu_{e}, \nu_{\mu }, \nu_{\tau}$  neutrinos are
passing through vacuum, they  will  be  converted  into
superpositions  of  the $\nu_{1}, \nu _{2}, \nu_{3}$  owing  to
the presence  of  the interactions violating  the  lepton number
of neutrinos and  will be left on  their mass   shells.  And,
then, oscillations of the $\nu_{e}, \nu_{\mu}, \nu_{\tau}$
neutrinos will  take  place according to the standard scheme [4].
In the case of two neutrino oscillations we will obtain
expressions equivalent to expressions (1)$\div$(8) and for the
case of three neutrino oscillations the common expression was
given in [14] for $V$ in all possible cases .
\par
Whether these oscillations are real or virtual, it will be
determined by the masses of the  physically observed neutrinos
$\nu_{e}, \nu_{\mu}, \nu_{\tau}$.
\par
i) If the masses of the $\nu_{e}, \nu_{\mu }, \nu_{\tau}$
neutrinos  are equal, then the real oscillation of the neutrinos
will take  place.
\par
ii) If  the masses  of  the $\nu_{e}, \nu _{\mu }, \nu _{\tau}$
are  not equal, then the virtual oscillation of  the  neutrinos
will  take place (the time of neutrino transitions will be defined
by uncertainty relation). To make these neutrinos real, these
neutrinos must participate in the quasielastic interactions, in
order to undergo transition to the mass shell of the other
appropriate neutrinos in analogy with $\gamma  - \rho ^{o}$
transition  in the vector meson dominance model. It is necessary
to take into account that in contrast to the strong interactions,
the dependence on squared transferring momentum in the weak
interactions has a flat form since $W$ boson has a huge mass. It
means that at weak interactions of oscillating neutrinos in matter
(detector) they transit on their mass shell and there an
additional dependence of squared transferring momentum does not
appear. In case ii) enhancement of neutrino oscillations will take
place if the mixing angle is small at neutrinos passing through  a
bulk of matter [15].
\par
So the neutrino mixings (oscillations) appear due to the fact that
at neutrino creating the eigenstates of the weak interactions the
$\nu_e, \nu_\mu, \nu_\tau$ neutrino states are produced but not
the eigenstates of the weak interaction violating lepton numbers
(i.e. $\nu_1, \nu_2, \nu_3$ neutrino states). And then when
neutrinos are passing through vacuum they are converted into
superpositions of $\nu_1, \nu_2, \nu_3$ neutrinos and through
these intermediate states they are converted from one type into
the other type. If $\nu_1, \nu_2, \nu_3$ neutrinos were originally
produced, then the mixings (oscillations) would not have taken
place since in the weak interaction where $\nu_e, \nu_\mu,
\nu_\tau$ neutrinos are produced the lepton numbers are conserved.
\par
In the case of three neutrino types the probability of $\nu_e \to
\nu_e$ transitions  has the following form [14]:
$$
P(\nu_e \to \nu_e, t)= 1 - cos^4(\beta)sin^2(2 \theta) sin^2(t
(E_1-E_2)/2) -
$$
$$
cos^2(\theta) sin^2(2 \beta) sin^2(t (E_1-E_3)/2) -
$$
$$
- sin^2(\theta) sin^2(2 \beta) sin^2(t (E_2-E_3)/2) , \eqno(22)
$$
where $E_1, E_2, E_3$ are energy of $\nu_1, \nu_2, \nu_3 \to x$
neutrinos and $E_x = \sqrt{p^2 + m^2_x}$.
\par
Since lengths of neutrino oscillations
$$
L_{i, j} = 2\pi  {p \over {\mid m^{2}_{2} - m^{2}_{1} \mid}} \quad
i \ne j = 1, 2, 3 , \eqno(23)
$$
are different, then the expression of probability for neutrino
oscillations at small distances has a simpler form. For example,
for $\nu_e \to \nu_e$ oscillations we have
\par
$$
P(\nu_e \rightarrow \nu_e) = 1 -  \sin^{2} 2\theta sin^2
((m^{2}_{2} - m^{2}_{1}) / 2p) t , \eqno(24)
$$
where
$$
sin^2 \theta= 1/2 - \frac{(m_{\nu_e} - m_{\nu_\mu})}{2
\sqrt{(m_{\nu_e} - m_{\nu_\mu})^2 +(2 m_{\nu_e \nu_\mu})^2}} ,
\eqno(25)
$$
and
$$
sin^2 2\theta = \frac{(2m_{\nu_{e} \nu_{\mu}})^2} {(m_{\nu_e} -
m_{\nu_\mu})^2 +(2m_{\nu_e \nu_{\mu}})^2} , \eqno(26)
$$
\par
It is interesting to remark that expression (26) can be obtained
from the Breit-Wigner distribution [16]
$$
P \sim \frac{(\Gamma/2)^2}{(E - E_0)^2 + (\Gamma/2)^2}   ,
\eqno(27)
$$
by using the following substitutions:
$$
E = m_{\nu_e},\hspace{0.2cm} E_0 = m_{\nu_\mu},\hspace{0.2cm}
\Gamma/2 = 2m_{\nu_e, \nu_\mu} , \eqno(28)
$$
where $\Gamma/2 \equiv W(... )$ is a width of $\nu_e
\leftrightarrow \nu_\mu$ transitions, i.e., virtual neutrino
oscillations keep in within the uncertainty relation. Then we can
interpret nondiagonal mass terms as widths of neutrino
transitions. In the general case these widths can be computed by
using a standard method [17].
\par
If $m_{\nu_e, \nu_\mu}$ differs from zero, then Exp. (26) gives a
probability of $\nu_e \leftrightarrow \nu_\mu$ transitions and
then the probability of $\nu_e \leftrightarrow \nu_\mu$
transitions is defined by these neutrino masses and width
(nondiagonal mass term) of their transitions. If $m_{\nu_e,
\nu_\mu} = 0$, then the ${\nu_e \leftrightarrow \nu_\mu}$
transitions are forbidden. So, this is a solution of the problem
of the origin of the mixing angle in the theory of vacuum
oscillations in the scheme of mass mixings.
\par
It is necessary to remark that in this corrected (alternative)
scheme of neutrino oscillations, in contrast to the standard
theory, oscillations of neutrinos with equal masses are real ones
and the oscillations of neutrinos with different masses are
virtual ones and then the problem of energy momentum conservation
as well as the problem of neutrino disintegrations as wave
packets, are solved.
\par
In the above considered scheme of neutrino oscillations at
neutrino oscillations their masses change (for example $m_{\nu_e}
\to m_{\nu_\mu}$). Theoretically neutrino transitions without
changing their masses are also possible [17]. In this case the
mixing angles are maximal ($\pi/4$). The author proposed another
mechanism (model) of neutrino transitions, which is generated by
charge (couple constant) mixings,  analogous to the model of
vector dominance, i.e., the model of $\gamma \to
\rho^o$ transitions [18]. \\

\par
{\bf 4. Conclusions} \\

\par
In the standard theory of neutrino oscillations it is supposed
that physical observed neutrino states $\nu_{e}, \nu_{\mu },
\nu_{\tau}$ have no definite masses, that they are initially
produced as a mixture of the $\nu_{1}, \nu_{2}, \nu_{3}$ neutrino
states (are produced as a wave packet), and that neutrino
oscillations are the real ones. Then this wave packet must
decompose at a definite distance into constituent parts and
neutrino oscillations must disappear. It has been shown that these
suppositions lead to violation of the law of energy and momentum
conservation. An alternative scheme of neutrino oscillations
obtained within the framework of particle physics has been
considered where the above mentioned shortcomings are absent, the
oscillations of neutrinos with equal masses are the real ones, and
the oscillations of neutrinos with different masses are virtual
ones. Expressions for probabilities of neutrino transitions
(oscillations) in the
alternative (corrected) scheme has been given. \\

\par
{\bf References}

\par
\noindent 1. Pontecorvo B. M., Soviet Journ. JETP, 1957, v. 33,
p.549;
\par
JETP, 1958,  v.34, p.247.
\par
\noindent 2. Maki Z. et al., Prog.Theor. Phys., 1962, vol.28,
p.870.
\par
\noindent 3. Pontecorvo B. M., Soviet Journ. JETP, 1967, v. 53,
p.1717.
\par
\noindent 4. Bilenky S.M., Pontecorvo B.M., Phys. Rep.,
C41(1978)225;
\par
Boehm F., Vogel P., Physics of Massive Neutrinos: Cambridge
\par
Univ. Press, 1987, p.27, p.121;
\par
Bilenky S.M., Petcov S.T., Rev. of Mod.  Phys., 1977, v.59, p.631.
\par
 Gribov V., Pontecorvo B.M., Phys. Lett. B, 1969, vol.28,
p.493.
\par
\noindent 5. Schiff L. I., Quantum Mechanics, McRam, ..., London,
1955.
\par
\noindent 6. Kayser B., Phys. Rev. D24, 1981, p.110.
\par
M. Zralik, Acta phys. Pol. B28, 1997, p.2225.
\par
\noindent 7. Schweber S., An Introduction to Relativistic Quantum
 Field Theory,
\par
 Row, ..., New York, 1961.
\par
 Ta-Pei Cheng, Ling-Fong Li, Gauge Theory of Elementary
Particle
\par Physics, Clarendon Press-Oxford, 1984.
\par
\noindent 8. Beshtoev Kh.M., Phys. of Elem. Part. and Atomic Nucl.
\par
(Particles and Nuclei), 1996, v.27, p.53.
\par
\noindent 9. Beshtoev Kh.M., JINR Commun. E2-92-318, Dubna, 1992;
\par
JINR Rapid Communications, N3[71]-95.
\par
\noindent 10. Kameda J., Proceedings of ICRC 2001, August 2001,
Germany,
\par
Hamburg, p.1057.
\par
Fukuda  S. et al,. Phys.   Rev. Lett., 2001, v.25, p.5651;
\par
Phys. Lett. B, 539, 2002,  p.179.
\par
\noindent 11. Ahmad Q. R. et al., Internet Pub. nucl-ex/0106015,
June 2001.
\par
Ahmad  Q. R. et al., Phys. Rev. Lett. 2002, v. 89, p.011301-1;
\par
Phys. Rev. Lett.  2002,v.  89, p.011302-1.
\par
\noindent 12. Beshtoev Kh.M., JINR Commun. E2-92-318, Dubna, 1992;
\par
JINR Rapid Communications, N3[71]-95; HEP-PH/9911513;
\par
 The Hadronic Journal, v.23, 2000, p.477;
\par
Proceedings of 27th Intern. Cosmic Ray Conf., Germany,
\par
Hamburg, 7-15 August 2001, v.3, p. 1186.
\par
\noindent 13. L. Maiani, Proc. Int. Symp.  on  Lepton-Photon
\par
Inter., Hamburg, DESY, p.867.
\par
\noindent 14. Beshtoev Kh.M., JINR Commun, E2-2006-16, Dubna,
2006.
\par
\noindent 15. Beshtoev Kh.M., JINR Commun, E2-93-297, Dubna, 1993;
\par
JINR Commun. E2-94-46; Hadronic Journal, 1995, vol 18, p.165.
\par
\noindent 16. Blatt J.M., Weiscopff V.F., The Theory of Nuclear
Reactions,
\par
INR T.R. 42.
\par
\noindent 17. Beshtoev Kh.M., HEP-PH/9911513;
\par
 The Hadronic Journal, v.23, 2000, p.477;
\par
Proceedings of 27th Intern. Cosmic Ray Conf., Germany,
\par
Hamburg, 7-15 August 2001, v.3, p. 1186.
\par
Beshtoev Kh.M., JINR Commun. E2-99-307, Dubna, 1999;
\par
JINR Commun. E2-99-306, Dubna, 1999.
\par
\noindent 18. Sakurai  J.J., Currents  and  Mesons,  The  Univ of
Chicago Press, 1967.
\par
Beshtoev Kh. M., JNR of USSR Academy Science P-217,
\par
Moscow, 1981.

\end{document}